\documentstyle[preprint,aps,epsfig]{revtex}

\def\be{\begin{equation}}
\def\ee{\end{equation}}
\def\ep{\epsilon}
\def\t{\tilde}
\def\sg{\sigma}
\def\et{\tilde{\epsilon}}

\begin{document}
\draft
\title{First-Principles Approach to Electrorotation Assay}
\author{J. P. Huang and K. W. Yu}
\address{Department of Physics, The Chinese University of Hong Kong,
 Shatin, NT, Hong Kong}
\maketitle

\begin{abstract}
We have presented a theoretical study of electrorotation assay based on 
the spectral representation theory. We consider unshelled and shelled 
spheroidal particles as an extension to spherical ones.
From the theoretical analysis, we find that the coating can change the 
characteristic frequency at which the maximum rotational angular velocity 
occurs.
The shift in the characteristic frequency is attributed to a change in the 
dielectric properties of the bead-coating complex with respect to those of 
the unshelled particles. 
By adjusting the dielectric properties and the thickness of the coating, 
it is possible to obtain good agreement between our theoretical 
predictions and the assay data.

\end{abstract}
\vskip 5mm \pacs{PACS Number(s): 82.70.-y, 87.22.Bt, 77.22.Gm, 77.84.Nh}

\section{Introduction}

When a suspension of colloidal particles or biological cells is exposed to 
an external electric field, the analysis of the frequency-dependent response
yields valuable information on various processes, like the structural 
(Maxwell-Wagner) polarization effects \cite{Gimsa,Gimsa99}.
The polarization is characterized by a variety of the characteristic 
frequency-dependent changes.
While the polarization of biological cells can be investigated by the method
of dielectric spectroscopy \cite{Asami80}, conventional dielectrophoresis
and electrorotation (ER) analyze the frequency dependence of translations 
and rotations of single cells in an inhomogeneous and rotating external
field, respectively \cite{Fuhr,Gimsa91}.
With the recent advent of experimental techniques such as automated video 
analysis \cite{Gasperis} as well as light scattering methods \cite{Gimsa99}, 
the cell movements can be accurately monitored.
In ER, the frequency dependence of the polarization leads to a phase shift 
between the induced dipole moment and the rotating field, giving rise to a
torque acting on the particle which causes the rotation of the individual 
cell.

The phenomenon of ER can be developed into a useful technique known as 
ER assay. The ER assay combines antibody technology with ER to detect 
analytes in aqueous solutions. 
The analyte to be detected is bound to a latex bead of known dielectric
properties to form the analyte-bead complex, which causes a change in the 
dielectric properties. 
The change can be detected by ER technique, thus allowing the rapid and 
accurate detection of analytes in aqueous solutions.
This method can be used to detect various analytes, the selection of 
which can be controlled by the proper choice of binding agents. 

In this work, we propose the use of the spectral representation 
\cite{Bergman} for analyzing the electrorotation of particles in 
suspensions.
The spectral representation is a rigorous mathematical formalism of 
the effective dielectric constant of a two-phase composite material 
\cite{Bergman}. 
It offers the advantage of the separation of material parameters 
(namely the dielectric constant and conductivity) from the cell
structure information, thus simplifying the study. 
From the spectral representation, one can readily derive the dielectric 
dispersion spectrum, with the dispersion strength as well as the 
characteristic frequency being explicitly expressed in terms of the 
structure parameters and the materials parameters of the cell suspension 
(see section II.B below). 
The actual shape of the real and imaginary parts of the permittivity
over the relaxation region can be uniquely determined by the Debye relaxation
spectrum, parametrized by the characteristic frequencies and the dispersion
strengths. So, we can study the impact of these parameters on the dispersion
spectrum directly. The same formalism has been used recently to study the 
dielectric behavior of cell suspensions \cite{Lei}.

In this connection, we mention alternative methods, namely, by solving 
the Laplace's equation directly and Gimsa's approach based on equivalent 
circuits \cite{Gimsa,Gimsa99}. To our knowledge, none of these methods 
separate microstructure parameters from material information.

\section{FORMALISM}

We regard a suspension as a composite system consisting of 
spherical or spheroidal particles of complex dielectric constant 
$\t{\ep}_1$ dispersed in a host medium of $\t{\ep}_2$.
A uniform electric field ${\bf E}_0=E_0 \hat{\bf z}$ is applied to the
composites along the $z$-axis. 
We briefly review the spectral representation theory of the effective 
dielectric constant to establish notations.

 \subsection{Spectral representation}

The spectral representation is a mathematical transformation of the
complex effective dielectric constant $\t{\ep}_e$. 
In its original form \cite{Bergman}, a two-phase composite material is 
considered, in which inclusions of complex dielectric constant 
$\t{\ep}_1$ and volume fraction $p$ are randomly embedded in a host 
medium of $\t{\ep}_2$.
The complex effective dielectric constant $\t{\ep}_e$ will in general 
depend on the constituent dielectric constants, the volume fraction 
of inclusions as well as the detailed microstructure of the 
composite materials.

The essence of the spectral representation is to define the following 
transformations. If we denote a complex material parameter
 \be
\t{s}=\left(1 - {\t{\ep}_1\over \t{\ep}_2} \right)^{-1},
 \ee
then the reduced effective dielectric constant
 \be
w(\t{s})=1 - {\t{\ep}_e\over \t{\ep}_2}, 
 \ee
can be written as
 \be
w(\t{s}) = \sum_n {F_n \over \t{s}-s_n},
\label{Fs}
 \ee
where $n$ is a positive integer, i.e., $n=1, 2, ...$, and $F_n$ and $s_n$, 
are the $n$-th microstructure parameters of the composite materials 
\cite{Bergman}. In Eq.(\ref{Fs}), $0\le s_n<1$ is a real number,
while $F_n$ satisfies a sum rule \cite{Bergman}:
 \be
\sum_n F_n=p,
 \ee
where $p$ is the volume fraction of the suspended cells.

As a result, the spectral representation is a useful theory which helps 
separate the material property from the geometric information. 
In what follows, we illustrate the spectral representation by the 
capacitance of simple geometry. 
In particular, a parallel-plate capacitor is considered as an example. 
We will discuss two cases, namely, the series combination and the 
parallel combination.

In the first case, if one inserts a dielectric slab of dielectric constant 
$\t{\ep}_1$ and thickness $h_1$, as well as a dielectric of $\t{\ep}_2$ and 
thickness $h_2$ (both of the same area $A$), 
into a parallel-plate capacitor of total thickness $h=h_1+h_2$, 
the overall capacitance $C$ is given by
$$
C^{-1}=C_1^{-1}+C_2^{-1}
$$
where $C_1=\t{\ep}_1 A/h_1$, and $C_2=\t{\ep}_2 A/h_2$, $A$ being the area of a 
plate. 
On the other hand, we may define the equivalent capacitance as 
$C=\t{\ep}_e A/h$, where $\t{\ep}_e$ is the effective dielectric constant.
That is, we replace the composite dielectric by a homogeneous dielectric
of dielectric constant $\t{\ep}_e$.

Let $\t{\ep}_1=\t{\ep}_2(1-1/\t{s})$, we can express $C$ in the spectral 
representation, 
$$
C=\frac{A \t{\ep}_2}{h}-\frac{A\t{\ep}_2 h_1/h^2}{\t{s}-h_2/h}.
$$
In accord with the spectral representation, one may introduce 
$w(\t{s})=1-\t{\ep}_e/\t{\ep}_2$, which is in fact the same as $w(\t{s})=1-C/C_0$, 
where $C_0$ is the capacitance when the plates are all filled with a 
dielectric material of $\t{\ep}_2$, namely, $C_0=\t{\ep}_2 A/h$. 
Thus we obtain
$$
w(\t{s})=\frac{h_1/h}{s-h_2/h}.
$$
from which we find that the material parameter is separated from the 
geometric parameter.
The comparison of $w(\t{s})$ with Eq.(\ref{Fs}) yields
 $$
 F_1=h_1/h,\ \ \ s_1=h_2/h.
 $$
We should remark that $F_1$ obtained herein is just equal to the 
volume fraction of the dielectric of $\t{\ep}_1$, and that $s_1$ satisfies 
$0\le s_1<1$, as required by the spectral representation theory.
  
Next we consider the parallel combination.
If one inserts a material of dielectric constant $\t{\ep}_1$ and area 
$w_1$ as well as a dielectric of $\t{\ep}_2$ and area $w_2$
(both of the same thickness $h$), 
into a parallel-plate capacitor of total area $A=w_1+w_2$, 
the overall capacitance $C$ is given by
$$
C=C_1+C_2
$$
where $C_1=\t{\ep}_1 w_1/h$, and $\t{\ep}_2=\t{\ep}_2 w_2/h$. Similarly, after
introducing the effective dielectric constant $\t{\ep}_e$, we may define 
the overall capacitance as $C=\t{\ep}_e A/h$.
   
Again, in the spectral representation, let $\t{\ep}_1=\t{\ep}_2(1-1/\t{s})$, then 
$$
C=\frac{\t{\ep}_2 A}{h}-\frac{\t{\ep}_2 w_1}{h \t{s}}.
$$
Writing $w(\t{s})=1-C/C_0$, we obtain
$$
w(\t{s})=\frac{w_1/A}{\t{s}}.
$$
From this equation, the material parameter is also found to be separated 
from the geometric parameter.
It is clear that $F_1=w_1/A$, i.e., the volume fraction of the 
dielectric of $\t{\ep}_1$, and $s_1=0$. 

 \subsection{Shell-spheroidal model}
 
For inclusions of arbitrary shape, the spectral representation can only be
solved numerically \cite{Bergman}. However, analytic solutions can be 
obtained for isolated spherical and ellipsoidal particles.
For dilute suspensions of prolate spheroidal particles, the particles 
can be regarded as noninteracting.
The problem is simplified to the calculation of $s_n$ with a single 
particle, which can be solved exactly.
 
In fact, the suspension of shell-spheroidal particles dispersed in a host 
medium is a three-component system. Although the spectral representation is 
generally valid for two-component composites, we have recently shown that 
it also applies to composites of shelled spheres randomly embedded in a host 
medium \cite{Yuen}.
Similarly, we will show that the spectral representation also applies to 
the suspension of spheroidal particles of complex dielectric constant 
$\t{\ep}_1$ coated with a shell of $\t{\ep}_s$ dispersed in a host 
medium of $\t{\ep}_2$, where
 \be
 \t{\ep}=\ep+\sigma/i2\pi f\nonumber
 \ee 
where $f$ is the frequency of the applied field. In what follows, we will 
show that from the spectral representation, one can obtain the analytic 
expressions for the characteristic frequency at which the maximum 
electrorotation velocity occurs.
The depolarization factors of the spheroidal particles will be described by 
a sum rule
 \be
 L_z+2L_{xy}=1\nonumber
 \ee 
where $0<L_z\le 1/3$ and $L_{xy}$ are the depolarization factors along the 
$z$- and $x$- (or $y$-) axes of the spheroidal particle, respectively. 
In fact, $L_z=L_{xy}=1/3$ just indicates a spherical particle.

The phenomenon of electrorotation is based on the interaction between a 
time-varying electric field ${\bf E}$ and the induced dipole moment 
${\bf M}$. The dipole moment of the particle arises from the induced 
charges that accumulate at the interface of the particle.
As the prolate spheroidal particles are easily oriented along their long 
axes by the rotating field, we consider the orientation in which
the long axis lies within the field plane \cite{Gimsa2001}.
In fact, the extension of our theory to deal with oblate spheroid is 
straightforward. In doing so, it suffices to consider $1/3<L_z<1$ and
the formlism will remain unchanged.

The angle between ${\bf M}$ and ${\bf E}$ is denoted by $\theta$, where 
$\theta=\omega \times$ time and $\omega=2\pi f$ is angular velocity of 
the rotating electric field. 
The torque acting on the particle is given by the vector cross product 
between the electric field and the dipole moment, so that only the 
imaginary part of the dipole moment contributes to the electrorotation 
response.
In the steady state, the frequency-dependent rotation speed $\Omega(f)$, 
which results from the balance between the torque and the viscous drag, 
is given by
   \begin{eqnarray}
\Omega(f) &=& -F(\ep_2, E,\eta)Im(\t{b}_z \langle\cos^2\theta \rangle
 + \t{b}_{xy} \langle\sin^2 \theta\rangle), \nonumber \\
  &=& -F(\ep_2, E,\eta)Im(\t{b}_z/2
 + \t{b}_{xy}/2), 
   \end{eqnarray}
where $E$ is the strength of the applied electric field and $F$ is a 
coefficient which is inversely proportional to the dynamic viscosity 
$\eta$ of the host medium. For spherical cells, $F=-\ep_2 E^2/2\eta$.
In Eq.(7), $Im(\cdots)$ indicates the imaginary parts of $(\cdots)$, and 
the angular brackets denote a time average. 
Since the angular velocity of the rotating field is much greater than the 
electrorotation angular velocity, i.e., $\omega \gg \Omega$, the time 
averages are just equal to 1/2. For a single shelled spheroidal particle, 
the dipole factor $\t{b}_z$ is given by \cite{Gao}
 \be
\t{b}_z=\frac{1}{3}\frac{(\t{\ep}_s-\t{\ep}_2)
 [\et_s+L_z(\et_1-\et_s)]+(\et_1-\et_s)y[\et_s+L_z(\et_2-\et_s)]}
 {(\et_s-\et_1)(\et_2-\et_s)yL_z(1-L_z)+[\et_s+(\et_1-\et_s)L_z]
 [\et_2+(\et_s-\et_2)L_z]}
 \ee 
where $y$ is the volume ratio of the core to the whole shelled spheroid, 
while $b_{xy}$ can be obtained by replacing $L_z$ with $L_{xy}$ in Eq.(8).
  
We are now in a position to represent $b_z$ and $b_{xy}$ in the spectral 
representation. Let $\et_1=\et_2(1-1/\t{s})$, and assume
$x=\et_s/\et_2$, we obtain
   \be
   \t{b}_z=N.P.+\frac{F_1}{\t{s}-s_1}
   \ee 
where $N.P.$ denotes the nonresonant part \cite{Yuen} which vanishes in the 
limit of unshelled spheroidal cells. In Eq.(9), the various quantities 
are given by
 \begin{eqnarray}
 s_1&=&-\frac{\beta}{\gamma},\\
 F_1&=&\frac{-x^2y}{\alpha \gamma},\\
 N.P.&=&\frac{-L_z+L_zy+x(-1+2L_z+y-2L_zy)+x^2(1-L_z-y+L_zy)}{\alpha}\\
\alpha&=&3L_z-3L_z^2-3L_zy+3L_z^2y+x(3-6L_z+6L_z^2+6L_zy-6L_z^2y)\nonumber\\
      & &x^2(3L_z-3L_z^2-3L_zy+3L_z^2y)\nonumber \\
\beta&=&-L_z+L_z^2+L_zy-L_z^2y+x(-L_z^2-L_zy+L_z^2y)\nonumber\\
\gamma&=&L_z-L_z^2-L_zy+L_z^2y+x(1-2L_z+2L_z^2+2L_zy-2L_z^2y)\nonumber\\
      & &+x^2(L_z-L_z^2-L_zy+L_z^2y)\nonumber
 \end{eqnarray}
 
Note that we have assumed $x$ to be a real number, which will be 
justified below. 
After substituting $\et=\ep+\sg/i2\pi f$ into Eq.[9], we rewrite $\t{b}_z$ 
after simple manipulations
 \be
 \t{b}_z=(N.P.+\frac{F_1}{s-s_1})+\frac{\delta\ep_1}{1+if/f_{c1}}
 \ee
with $s=(1-\ep_1/\ep_2)^{-1}$ and $t=(1-\sg_1/\sg_2)^{-1}$, where
 \begin{eqnarray}
 \delta\ep_1&=&F_1\frac{s-t}{(t-s_1)(s-s_1)},\nonumber\\
f_{c1}&=&\frac{1}{2\pi}\frac{\sg_2}{\ep_2}\frac{s(t-s_1)}{t(s-s_1)}.\nonumber
 \end{eqnarray}
Similarly, we may rewrite $\t{b}_{xy}$ as 
 \be
 \t{b}_{xy}=(N.P.'+\frac{F_2}{s-s_2})+\frac{\delta\ep_2}{1+if/f_{c2}}.
 \ee 
 Therefore, we obtain
 
 \begin{eqnarray}
 \delta\ep_2&=&F_2\frac{s-t}{(t-s_2)(s-s_2)},\nonumber\\
 f_{c2}&=&\frac{1}{2\pi}\frac{\sg_2}{\ep_2}\frac{s(t-s_2)}{t(s-s_2)}.\nonumber
 \end{eqnarray}
Note that $N.P.'$, $s_2$ and $F_2$ are obtained by replacing $L_z$ with 
$L_{xy}$ in the expressions for $N.P.$, $s_1$ and $F_1$, respectively.

We have thus predicted that two characteristic frequencies may appear for 
unshelled or shelled spheroidal particles.
Previous theories were often limited to spherical particles,
i.e., $L_z=L_{xy}=1/3$. 
Therefore, only one characteristic frequency exists.
We should remark that even though two characteristic frequencies are 
predicted, only one of them is dominant (see below).
 
 \section{numerical calculations}

The model put forward in the previous section applies to various 
situations such as biological cells and polystyrene beads, etc. 
Here we perform numerical calculations to investigate the characteristic 
frequency. Let $s=1.1$, $t=-0.005$, and $\ep_2=80\ep_0$, where $\ep_0$ is 
the dielectric constant of the vacuum.
In Fig.1, we investigate the effect of particle shape on $s_1$ and 
$f_{c1}$ (upper panels), and $s_2$ and $f_{c2}$ (lower panels), for $z=2$ 
and $\sg_2=2.9 \times 10^{-5} Sm^{-1}$, where
$z=1/y$, and $z^{1/3}>1$ reflects the thickness 
of the shell (or coating). As is evident from the figure, an increase
in the dielectric constant ratio $x$ leads to a red-shift of the
characteristic frequency. 
For a certain $x$, a small depolarization factor, i.e., the particle is 
largely deviated from spherical shape, may yield a red-shift too.

In Fig.2, we investigate the effect of the depolarization factor on the 
quantities $-Im[b_z]/2$ and $-Im[b_z/2+b_{xy}/2]$. 
It is evident that the effect of $b_{xy}$ on the peak is small as $L_z$ is 
small, whereas it gives a large effect as $L_z$ is large. 
Generally speaking, the dipole moment along the x-(or y-) axis strongly 
affects both the location and the magnitude of the peak of rotation speed.
For spheroidal particles, only one peak is found, and the other peak 
predicted by the theory may be too small to observe.

In order to validate our theory, here we considered the spherical particles 
as a limiting case of our model.
In Fig.3, $s_1$ and $f_c$ are plotted versus $x$ for a spherical bead 
($L_z=1/3$), for different $z$ as $\sg_2=2.9\times 10^{-5} Sm^{-1}$. 
For large $x$, the shell thickness has only a minor effect on the 
characteristic frequency.
Moreover, a thick shell leads to a red shift (blue shift) of the 
characteristic frequency appearing when $x>1$ ($x<1$).
Also, all the $f_c$ predicted by different $z$ are smaller (larger) than 
that predicted one at $z=1$ (i.e., unshelled bead) for $x>1$ ($x<1$).

At $x=1$, i.e., the shell has the same dielectric constant as the host, 
all the characteristic frequencies predicted by different thickness of 
shell are the same. Similar conclusions can be obtained not only for 
spherical shape, but also for prolate spheroidal shape 
(not shown here).

We attempt to fit our theoretical predictions with experimental data,
which are extracted from an assay \cite{Burt}. 
In this assay, three cases were studied, all dealing with spherical 
particles: that is, unshelled beads, 
beads coated with an antibody with specificity for Giardia (Shell 1),
and beads coated with an antibody with specificity for Cryptosporidium 
(Shell 2).
The bead diameter is $6 \mu m$ according to Burt et al \cite{Burt}, 
while Giardia and Cryptosporidium are both $0.8\mu m$ in diameter. 
Let $\sg_2=2.18 \times 10^{-5} Sm^{-1}$, $z=6$, $x=7.06$ (Shell 1) and 
$4.63$ (Shell 2), and $F=0.353$ (Without shell), $1.629$ (Shell 1) and 
$2.278$(Shell 2). Good agreement between our theoretical predictions 
and the assay data is shown in Fig.4.
From our theory, it is easy to find the corresponding characteristic 
frequencies at which maximum rotational angular velocity occurs, 
$f_c=4.755\times 10^5Hz$ (Without shell), $10^5Hz$ (Shell 1) and 
$1.415\times 10^5Hz$ (Shell 2).
From this, we find that the coating leads to a red shift of the 
characteristic frequency.
It is because the dielectric properties of the bead-coating complex have 
been changed. 

We used $x>1$ in our fitting, i.e., $x=7.06$ and 4.63 for bead coated 
with an antibody with specificity for Giardia and Cryptosporidium, 
respectively. 
It is known that, for biological cells, such 
as Giardia and Cryptosporidium, they have structures such as the cell 
wall, plasma membrane and the cytoplasm, among which the cell wall has 
a larger conductivity than the suspending medium. Hence, we may safely 
take $x>1$. 

\section{Discussion and conclusion}

Here we would like to add some comments.
We would like to clarify the assumptions with our model in more detail. 
In fact, there is only one (rather than two) peak for each polarization 
in our theory. It is because we assumed the ratio of the shell to host 
dielectric constant $x$ to be a real and positive number. 
If we had retained the (indeed small) imaginary part of $x$ in our 
calculation, then we would have two peaks for each polarization. 
The conductivity-dominated peak would have occured at substantially lower
frequency. Thus, the neglect of the imaginary part of $x$ is to drop the
lower frequency peak.
Moreover, according to our calculations, there is one dominant pole 
associated with $b_z$ and two (degenerate) sub-dominant poles associated 
with $b_{xy}$ in the spectral representation. 
Thus, for shelled particles in the present work, only one peak has been 
shown.

We developed simple equations to describe the electrorotation of 
particles in a suspension from the spectral representation. 
These equations serve as a basis which describe the parameter dependence 
of the polarization and thereby enhances the applicability of various 
cell models for the analysis of the polarization mechanisms. 
In this connection, the shell-spheroidal cell model may readily be 
extended to multi-shell cell model.
However, we believe that the multi-shell nature of the cell may have 
a minor effect on the electrorotation spectrum.

We have considered the isolated cell case, which is a valid assumption
for low concentration of cells. However, for a higher concentration of 
cells, we should consider the mutual interaction between cells. For a
randomly dispersed cells suspension, we may replace the dielectric constant
of the host medium by the effective dielectric constant of the whole 
suspension.

When a strong rotating electric field is applied to a suspension,
the induced dipole moment will induce an overall attractive force between 
the polarized cells, leading to rapid formation of sheet-like structures 
in the plane of the rotating field. In reality there is a phase shift 
between the induced dipole moment of the structure and the applied field, 
and this can lead to electrorotation. However, the situation will be much 
more difficult than the single-cell case that we have studied because 
the many-body as well as multipolar interactions between the particles 
will produce a complicated electrorotation spectrum. 
Fortunately, our recently developed integral equation formalism \cite{Yu2000} 
can definitely help to solve for the Maxwell-Wagner relaxation spectrum.

In summary,
we have presented a theoretical study of electrorotation assay based on 
the spectral representation theory. We consider unshelled and shelled 
spheroidal particles as an extension to spherical ones.
From the theoretical analysis, we find that the coating can change the 
characteristic frequency at which maximum rotational angular velocity occurs.
By adjusting the dielectric properties and the thickness of the coating, 
it is possible to obtain good agreement between our theoretical 
predictions and the assay data.

\section*{Acknowledgments}
This work was supported by the Research Grants Council of the Hong Kong 
SAR Government under grant CUHK 4245/01P. 
J. P. H. is grateful to Dr. L. Gao and Dr. C. Xu for fruitful discussion.
K. W. Y. acknowledges the hospitality of Prof. Hong Sun when he visited 
the University of California at Berkeley and useful discussion with Prof. 
G. Q. Gu.

\begin{figure}[h]
\caption{Upper panels: $s1$ and $f_{c1}$ plotted against $x$ for 
different $L_z$ at $z=2$.
Lower panels: $s2$ and $f_{c2}$ plotted against $x$ for different 
$L_z$ at $z=2$.}
\end{figure}

\begin{figure}[h]
\caption{$-$Im$(b_z/2+b_{xy}/2)$ and $-$Im$b_z/2$ are plotted against
frequency $\omega$ for different $L_z$ at $z=6$, $x=2$ and 
$\sigma_2=2.9 \times 10^{-5}Sm^{-1}$.}
\end{figure}

\begin{figure}[h]
\caption{$s1$ and $f_c$ are plotted versus $x$ for different $z$ as 
$L_z=1/3$ (i.e., spherical shape).}
\end{figure}

\begin{figure}[h]
\caption{Curve fitting for $L_z=1/3$ (i.e., spherical particles).}
\end{figure}

\newpage
\centerline{\epsfig{file=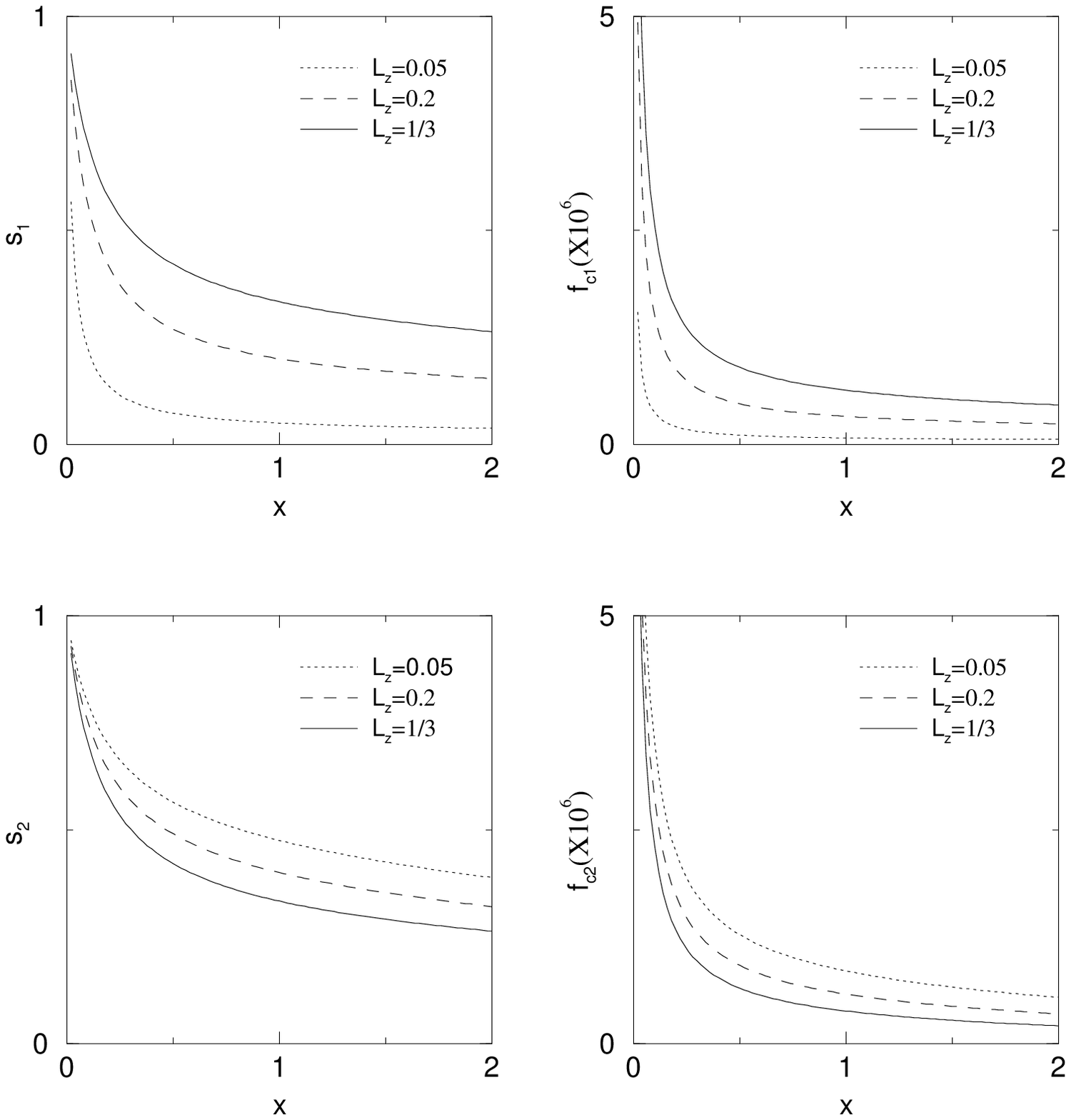,width=\linewidth}}
\centerline{Fig.1}

\newpage
\centerline{\epsfig{file=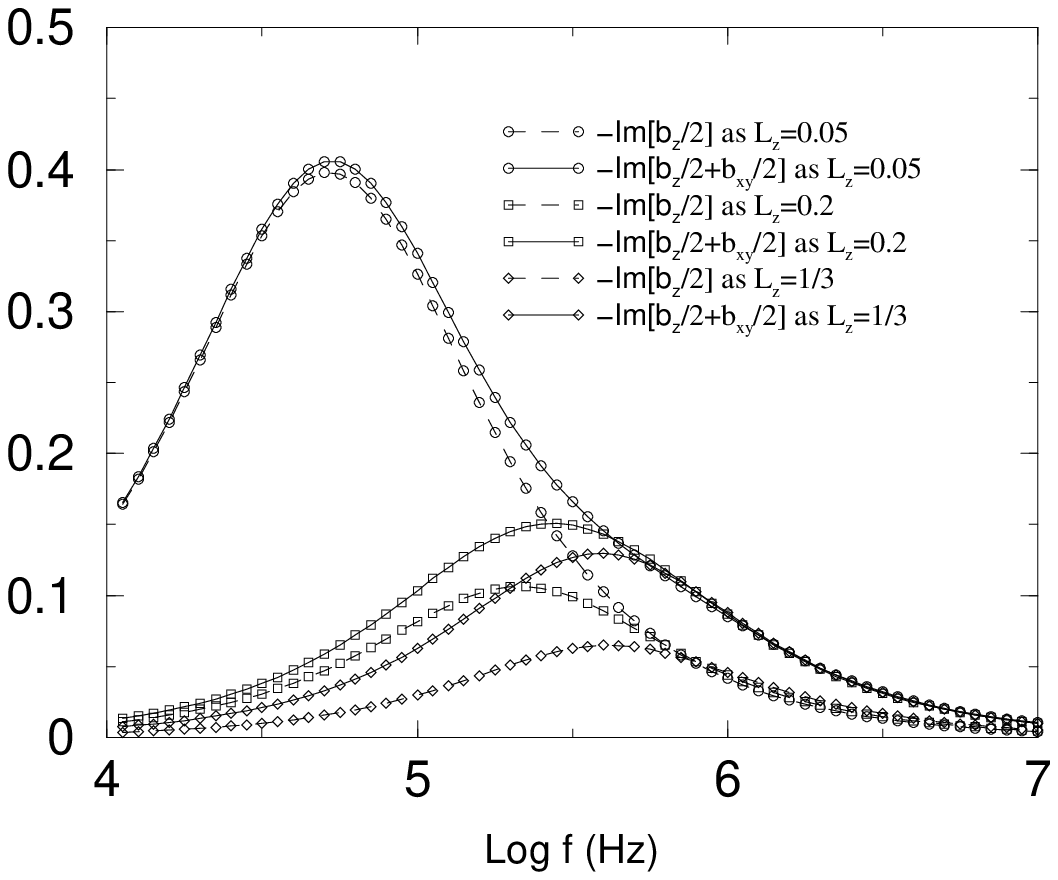,width=\linewidth}}
\centerline{Fig.2}

\newpage
\centerline{\epsfig{file=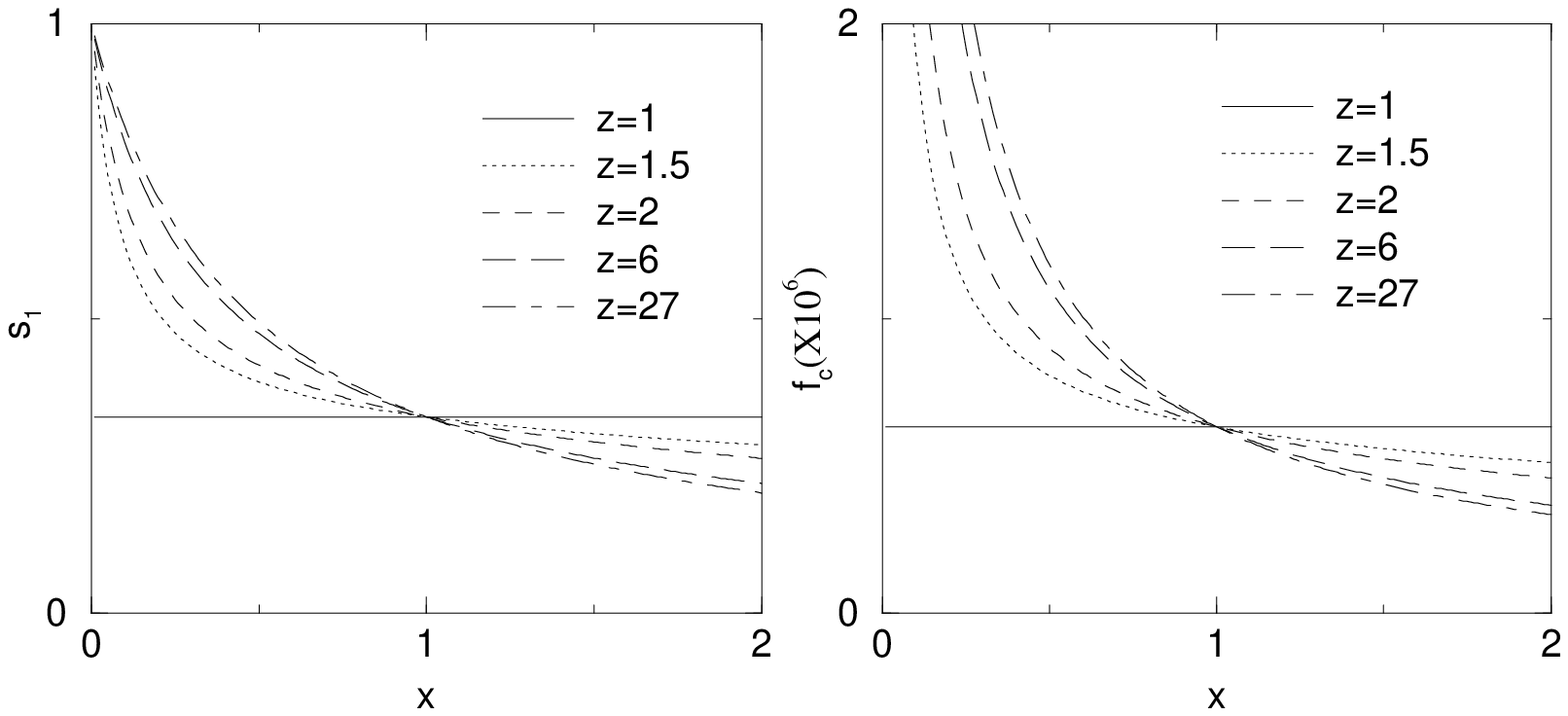,width=\linewidth}}
\centerline{Fig.3}

\newpage
\centerline{\epsfig{file=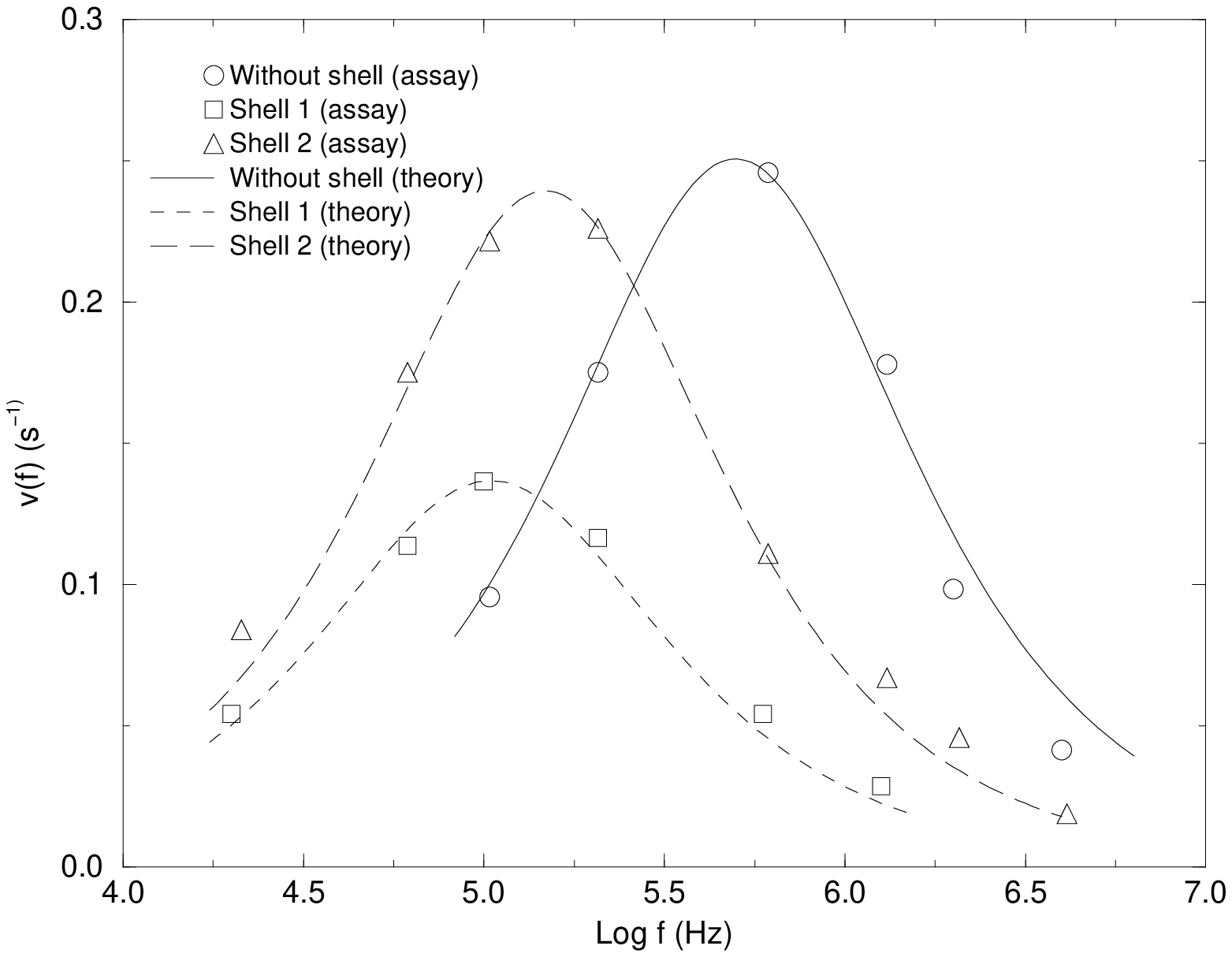,width=\linewidth}}
\centerline{Fig.4}


\begin{references}

\bibitem{Gimsa} For a review, see J. Gimsa and D. Wachner,
 Biophys. J. {\bf 77}, 1316 (1999).
\bibitem{Gimsa99} J. Gimsa, Ann. NY Acad. Sci. {\bf 873}, 287 (1999).
\bibitem{Asami80} K. Asami, T. Hanai and N. Koizumi, Jpn. J. Appl. Phys.
 {\bf 19}, 359 (1980).
\bibitem{Fuhr} G. Fuhr, J. Gimsa and R. Glaser, Stud. Biophys. {\bf 108},
 149 (1985).
\bibitem{Gimsa91} J. Gimsa, P. Marszalek, U. Lowe and T. Y. Tsong, 
 Biophys. J. {\bf 73}, 3309 (1991).
\bibitem{Gasperis} G. De Gasperis, X.-B. Wang, J. Yang, F. F. Becker 
 and P. R. C. Gascoyne, Meas. Sci. Technol. {\bf 9}, 518 (1998).

\bibitem{Bergman} D. J. Bergman, Phys. Rep. {\bf 43}, 379 (1978).  
\bibitem{Lei} J. Lei, Jones T. K. Wan, K. W. Yu and Hong Sun,
 Phys. Rev. E {\bf 64}, 012903 (2001).
\bibitem{Yuen} K. P. Yuen and K. W. Yu, J. Phys.: Condens. Matter 
 {\bf 9}, 4669 (1997).
\bibitem{Gimsa2001} It is not always necessary that prolate particles
 orient with their longest axis in the field plane. 
 Depending on frequency, object and medium properties also a perpendicular 
 orientation can be observed, see e.g.,
 J. Gimsa, Bioelectrochemistry {\bf 54}, 23 (2001).
\bibitem{Gao} L. Gao, Jones T. K. Wan, K. W. Yu and Z. Y. Li, J. Phys.: 
Condens. Matter {\bf 12}, 6825 (2000).
\bibitem{Burt} J. P. H.  Burt, K. L. Chan, D. Dawson, A. Parton and 
 R. Pethig, Ann. Biol. Clin. {\bf 54}, 253 (1996); see also 
 http://www.ibmm.informatics.bangor.ac.uk/pages/science/rot.htm
 for the basic science of electrorotation.

\bibitem{Yu2000} K. W. Yu, Hong Sun and Jones T. K. Wan, 
 Physica B {\bf 279}, 78 (2000).
\end{references}
\end{document}